\documentclass[superscriptaddress,twocolumn,showpacs,prb]{revtex4-1}
\usepackage[utf8]{inputenc} 
\usepackage{mathtools}
\usepackage{amsmath}
\usepackage{braket}
\usepackage{graphicx}
\usepackage{pgfplots}
\pgfplotsset{compat=1.14}
\usepackage{csquotes}
\usepackage{hhline}
\usepackage{tabularx}
\usepackage{subcaption}
\usepackage{amssymb}

\usepackage{dcolumn}
\usepackage{tabularx}
\usepackage{braket}
\usepackage{float}
\usepackage{listings}
\usepackage{color}
\usepackage{subcaption}
\usepackage{placeins}
\definecolor{mygreen}{rgb}{0,0.6,0}
\definecolor{mygray}{rgb}{0.5,0.5,0.5}
\definecolor{mymauve}{rgb}{0.58,0,0.82}

\newcolumntype{C}{>{\centering\arraybackslash}X}

\begin{document}

\title{Experimental Demonstration of Force Driven Quantum Harmonic Oscillator in IBM Quantum Computer}

\author{Alakesh Baishya}
\email{alakeshnit16@gmail.com}
\affiliation{Department of Physics,\\ National Institute of Technology, Silchar, 788010, India}
\author{Lingraj Kumar}
\email{lkumar1@student.nitw.ac.in}
\affiliation{Department of Physics,\\ National Institute of Technology, Warangal, 506004, Telengana, India}
\author{Bikash K. Behera}
\email{bkb18rs025@iiserkol.ac.in}
\author{Prasanta K. Panigrahi}
\email{pprasanta@iiserkol.ac.in}
\affiliation{Department of Physical Sciences,\\ Indian Institute of Science Education and Research Kolkata, Mohanpur 741246, West Bengal, India}

\begin{abstract}
Though algorithms for quantum simulation of Quantum Harmonic Oscillator (QHO) have been proposed, still they have not yet been experimentally verified. Here, for the first time, we demonstrate a quantum simulation of QHO in the presence of both time-varying and constant force field for both one and two dimensional case. New quantum circuits are developed to simulate both the one and two-dimensional QHO and are implemented on the real quantum chip ``ibmqx4". Experimental data, clearly illustrating the dynamics of QHO in the presence of time-dependent force field, are presented in graphs for different frequency parameters in the Hamiltonian picture.  
\end{abstract}
 

\maketitle
\section{Introduction}
Quantum simulation is one of the tremendously growing areas in the field of quantum computation which has significant goals and opportunities \cite{qho_CiracNatPhys2012}. From the past decades, this powerful area has been applied to variety of scientific disciplines, e.g., physics \cite{qho_NatPhys2012,qho_GeorgescuRMP2014,qho_SimonNat2011,qho_KimNat2010,qho_StruckScience2011}, quantum chemistry \cite{qho_CollessPRX2018,qho_McArdlearXiv2018}, quantum biology \cite{qho_LambertNatPhys2013,qho_MohantaarXiv2018}, and computer science \cite{qho_DiazARCP2006} to name a few. Several time-dependent mass harmonic oscillators including the most famous so-called Caldirola-Kanai oscillator \cite{qho_CaldirolaINC1941,qho_KanaiPTP1950} have been extensively studied over the past years \cite{qho_DaneshmandIJTP2017,qho_DaneshmandLP2016,qho_KimJPA2003,qho_Man'koJRLR}. IBM quantum experience, has played a considerable role from the recent years, the platform using which a number of research works have been performed in the field of quantum simulation. These include observation of Uhlmann phase \cite{qho_ViyuelanpjQI2018}, chemical isomerization reaction \cite{qho_HalderarXiv2018}, simulation of far-from-equilibrium dynamics \cite{qho_ZhukovQIP2018}, Ising model simulation \cite{qho_Cervera-Lierta2018}, quantum multi-particle tunneling \cite{qho_MalikRG2019}, quantum scrambling \cite{qho_AggarwalarXiv2018}, and simulation of Klein-Gordon equation \cite{qho_KapilarXiv2018} to name a few.  Other sub-disciplines such as developing quantum algorithms \cite{qho_GarciaJAMP2018,qho_RounakarXiv2018,qho_SisodiaQIP2017,qho_GangopadhyayQIP2018,qho_DeffnerHel2017,qho_YalcinkayaPRA2017,qho_5BKB6arXiv2018,qho_DasharXiv2018}, testing of quantum information theoretical tasks \cite{qho_VishnuQIP2018,qho_Alakesh2019,qho_HuffmanPRA2017,qho_AlsinaPRA2016,qho_KalraarXiv2017}, quantum cryptography \cite{qho_BeheraQIP2017,qho_Plesa2018,qho_MajumderarXiv2017,qho_SarkarRG2019}, quantum error correction \cite{qho_GhoshQIP2018,qho_Roffe2018,qho_SatyajitQIP2018,qho_HarperarXiv2018}, quantum applications \cite{qho_SchuldEPL2017,qho_2BKB6arXiv2018,qho_BKB6arXiv2017,qho_Solano2arXiv2017,qho_BeheraQIP2019,qho_BeheraarXiv2018} have also been explored. 

Quantum harmonic oscillator (QHO) is one of the most practiced models in the branch of physics. A large number of works in QHO have been developed, e.g., in 1989, q-analogue \cite{qho_MacfarlaneJPA1989} of it has been explored. Brownian motion has been repeatedly studied \cite{qho_AgarwalPR1971,qho_DaviesCMP1972,qho_FordJMP1964,qho_SchwingerJMP1961,qho_LindbladRMP1976}. A quantum theory for time-dependent harmonic oscillator has been proposed by Lewis and Riesenfeld \cite{qho_LewisJMP1969}. QHO has also been realized using continuous-variable quantum computation \cite{qho_BartlettIEEE2001,qho_DengSciRep2016}. As harmonic oscillators govern most of the dynamics of the physical systems, the analysis of its behavioral dynamics plays a significant role in physics. However, it becomes difficult to study its transient behavior in the presence of a time-varying force field. Most of the experiments that have direct experimental observation with QHO have been simulated using NMR quantum system \cite{qho_SomarooPRL1999,qho_HolsteinPR1940}. 

Unitary dilation of the contraction semigroup governing the dynamics of the system in case of QHO was studied \cite{qho_LatimerIOP2005}. Moreover, it has been observed that the energy of a QHO with a generically time-dependent but cyclic frequency $\omega(t_0)=\omega(0)$, which cannot decrease on an average if the system is originally in a stationary state after the system goes through a full cycle. The energy exchange always takes place in the direction from the macroscopic system (environment) to the quantum microscopic system \cite{qho_KonishiIJMP2006}. The harmonic oscillator is also designed by different processes, e.g., transistor harmonic oscillator \cite{qho_SFENCE1973}, adelic harmonic oscillator \cite{qho_Branko1995} etc.

It can be mentioned though there are quantum algorithms for the harmonic oscillator, till now no literature has developed the circuit implementation of the quantum harmonic oscillator. Here, we demonstrate the circuit implementation of the quantum harmonic oscillator and implement the quantum circuits on the real quantum chips provided by IBM Q Experience Beta version. We use the quantum chip ``ibmqx4" for designing the quantum circuits for the case of one and the two-dimensional harmonic oscillator. We obtain the experimental results and present the dynamics of the quantum harmonic oscillator through line graphs while considering different frequency parameters for both the cases.       

The paper is organized as follows. In Section \ref{qho_SecII}, the scheme of harmonic oscillator has been explained. Section \ref{qho_SecIII} discusses the procedure for the simulation of the quantum circuit and its experimental demonstration on the quantum computer. In Section \ref{qho_SecV}, we present the experimental results obtained after running the quantum circuits on the quantum chips. Finally, we conclude in Section \ref{qho_conclusion} with a discussion on the future directions of the work.   

\section{Scheme of Harmonic Oscillator \label{qho_SecII}}

Let's consider a closed system where a driven force field of a single frequency, $\omega$ is applied. It can be mentioned that the frequency of the driving field changes for a different set of parameters. The Hamiltonian of the force driven harmonic oscillator is given by,

\begin{equation}
\centering
\mathcal{H}=\hbar \omega_0(a^{\dagger}a+\frac{1}{2})+\hbar F(t)(a+a^{\dagger})
\label{qho_Eq1}
\end{equation}
Here, $\omega_0$ is the frequency of the oscillator, $a^{\dagger}$ is the creation operator, and a is the annihilation operator of energy quanta of the oscillator. Again, ${F(t)}= \frac{A cos(\omega t + \phi)}{\sqrt{2m\hbar}}$ is the periodic driving force, where, A is the amplitude, $\omega$ is the oscillation frequency of the periodic force, m is the mass of the oscillator and $\phi$ is the phase of the driving force, the constant $\hbar$ is set to 1 throughout the manuscript. For the experimental demonstration, we have taken two states of QHO, i.e., $\ket{0}$ as the ground state and $\ket{1}$ as the 1st excited state. The matrix form for the Hamiltonian of the given system is calculated to be,

\begin{equation}
   H=
  \left[ {\begin{array}{cc}
   \frac{1}{2} & - F(t) \\
    F(t)& \frac{3}{2} \\
  \end{array} } \right]
\label{qho_Eq2}
\end{equation}

In this case, the matrix form of a and $a^{\dagger}$ are taken as, $a=[0, 1;0, 0]$ and $a^{\dagger}=[0, 0;1, 0]$. Again, for simulating the behavior of the particle for a two-qubit system, we have chosen four states; ground state $\ket{00}$, first excited state $\ket{01}$, second excited state $\ket{10}$ and third excited $\ket{11}$. To construct the Hamiltonian, we consider annihilation operator as $(I\otimes X)_{12}(ACNOT)_{21}$ and creation operator as $(I\otimes X)_{12}(CNOT)_{21}$. Here, $O_{ij}$ means O operation is applied from $i^{th}$ qubit to $j^{th}$ qubit, where $i^{th}$ qubit acts as the control qubit and $j^{th}$ one acts as the target qubit. $ACNOT$ denotes here the anti-controlled operation where the $NOT$ gate is applied only when the control qubit is in $|0\rangle$ state. Whereas, in case of $CNOT$ operation, $NOT$ gate is applied on the target qubit only when the control qubit is in the $|1\rangle$ state. The matrix form for the two-qubit Hamiltonian is given as, $H=H_1 + H_2$,

\begin{equation}
   H=
  \frac{3}{2}\hbar\omega_0\left[ {\begin{array}{cccc}
   1 & 0 & 0 & 0 \\
    0 & 1 & 0 & 0\\
    0 & 0 & 1 & 0\\
    0 & 0 & 0 & 1\\
  \end{array} } \right] + \hbar tF(t)\left[ {\begin{array}{cccc}
   0 & 1 & 0 & 1 \\
    1 & 0 & 1 & 0\\
    0 & 1 & 0 & 1\\
    1 & 0 & 1 & 0\\
  \end{array} } \right] 
  \label{qho_Eq3}
\end{equation}

\section{Quantum circuit implementation for demonstrating the quantum harmonic oscillator \label{qho_SecIII}}
\begin{figure}[]
\includegraphics[scale=0.6]{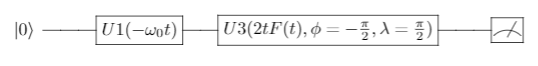}
\caption{Quantum circuit illustrating the force driven time dependent quantum harmonic oscillator for the single qubit system.}
\label{qho_Fig1}
\end{figure}

For simulating the Hamiltonian of both the single-qubit and two-qubit system, we have used first order trotter decomposition,
\begin{equation}
e^{-i\mathcal{H} t} = e^{-i \mathcal{H}_1t}e^{-i \mathcal{H}_2 t}...e^{-i \mathcal{H}_n t} +O\big(t^2)
\label{qho_Eq4}
\end{equation}

where, $\mathcal{H}=\mathcal{H}_1 +\mathcal{H}_2+....+\mathcal{H}_n$. Here the system Hamiltonian for the single qubit case given in Eq. \eqref{qho_Eq2} can be written in the form $\mathcal{H}=\omega_0 I-\frac{\omega_0}{2} Z+F(t)X$. The first term could be neglected because it is a constant times an identity matrix, whose corresponding unitary operator itself is an identity operation with some global phase, $e^{-i\omega_0}t$, hence can be neglected in the simulation part. Then the total unitary operator, $U_{single}$ of the single-qubit Hamiltonian can be given as, $U_{single}=U_1.U_2$, where $U_1=e^{-i\mathcal{H}_2 t}$ and $U_2=e^{-i\mathcal{H}_3 t}$. The form of $U_1$ and $U_2$ can be written as, $U_1 = [1, 0;0, e^{-i\omega_0 t}]$, and $U_2= [cos(tF(t)), -isin(tF(t)); -isin(tF(t)), cos(tF(t))]$. $U_1$ is implemented on the quantum computer by using the $U1(\theta)$ gate, whose form is given as, $U1(\theta)=[1, 0; 0, e^{i\theta}]$. The value of $\theta$ is taken to be $-\omega_0 t$ for designing $U_1$ unitary operator. Similarly, $U_2$ can be prepared with the $U3(\theta, \phi, \lambda)$ gate whose form is given as, $U3(\theta, \phi, \lambda)=[\cos{\frac{\theta}{2}}, e^{-i\lambda}\sin{\frac{\theta}{2}}; e^{i\phi}\sin{\frac{\theta}{2}}, e^{i(\lambda+\phi)}\cos{\frac{\theta}{2}}]$. Here, $\theta$, $\phi$ and $\lambda$ are chosen to be $2tF(t)$, $-\pi/2$ and $\pi/2$ respectively. It can be observed from the Fig. \ref{qho_Fig1} that $U1$ and $U3$ gates with proper parameters are applied on the system qubit initially prepared in the $|0\rangle$ state.  

From the two-qubit Hamiltonian (Eq. \eqref{qho_Eq3}), it can be observed that the first term is a scalar multiplication of the identity matrix, which again has no contribution in simulating the Hamiltonian. Hence we only consider the second term, which has an actual contribution in the state evolution of the two-qubit system. The unitary operator corresponding to that term is given as follows;      

\begin{equation}
  U = \left[ {\begin{array}{cccc}
   1 & -itF(t) & 0 & -itF(t) \\
    -itF(t) & 1 & itF(t) & 0\\
    0 & -itF(t) & 1 & -itF(t)\\
    -itF(t)& 0 & -itF(t) & 1\\
  \end{array} } \right] 
\label{qho_Eq5}
\end{equation}
The above unitary operator is obtained by expanding only first two terms of the exponential $e^{-i\mathcal{H}t}$ and neglecting higher order terms. Here we make the following assumptions to simulate the above operator as the unitary operator; We have taken the assumptions as $\cos{\theta}$ $\approx$ 1, $\sin{\theta} \approx tF(t)$. 

\begin{figure*}
\includegraphics[width=\linewidth]{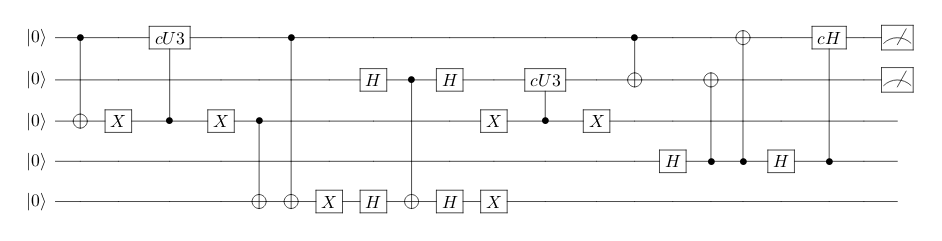}
\caption{Quantum Circuit to demonstrate the force driven time-dependent quantum harmonic oscillator for the two-qubit system. The first two qubits represent the system, and the last three qubits are the ancilla qubits used to apply the Hamiltonian on the system.}
\label{qho_Fig2}
\end{figure*}
From Fig. \ref{qho_Fig2}, it can be seen that in total, five qubits are used to simulate the two-qubit Hamiltonian. The first two qubits are used to represent the system, and the last three qubits are the ancilla qubits used to properly apply the Hamiltonian on the two-qubit system. All the ancillary qubits are initially prepared in $|0\rangle$ state. As observed from Fig. \ref{qho_Fig2}, a series of the following operations $CNOT_{13}$, $AC-U3_{31}$, $CCNOT_{132}$, $AC-U3_{32}$, $CNOT_{12}$, $H_4$, $CNOT_{42}$, $CNOT_{41}$, $H_{4}$, $CH_{41}$ are used in the quantum circuit to construct the unitary operator for the Hamiltonian. Here single-qubit gate, $O_{i}$ means $O$ operation is applied on the $i^{th}$ qubit, for the two-qubit operation, $O_{ij}$ denote that $O$ operation acts on both the $i^{th}$ and $j^{th}$ qubit where $i$ acts as the control and $j$ acts as the target qubit. Similarly, for the three-qubit operation, $O_{ijk}$ denote that $O$ operation acts on the three qubits i, j and k, where i and j act as the control qubit and k acts as the target qubit. In the above series of gates, $AC$ stands for Anti-Controlled operation, and $CC$ stands for Controlled-Controlled operation. In the above paragraph, the form of the $U3$ gate can be found. The parameters of $U3$ are taken as $\theta$=0 to 0.5, $\phi=-\pi/2$ and $\lambda =\pi/2$. The form of $NOT$ and Hadamard gate can be given as; $X=[0, 1; 1, 0]$, $H=\frac{1}{\sqrt{2}}[1, 1; 1, -1]$ respectively. The measurements are performed only on the system qubit to study its dynamics over a time steps and for different frequency parameters. 

\section{Results \label{qho_SecV}} 
\begin{figure*}[!ht]
    \centering
     \begin{subfigure}{.5\textwidth}
    \includegraphics[width=\linewidth]{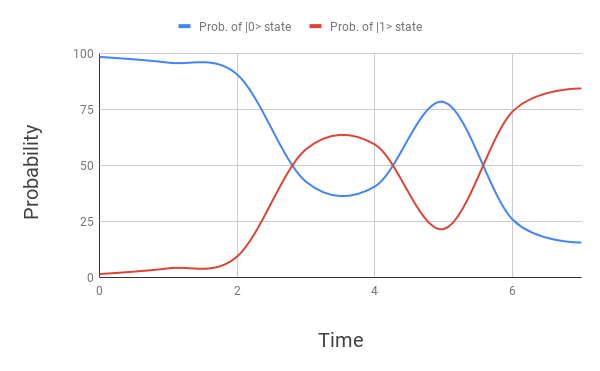}
    \caption{ For $\omega$=1}
    \label{qho_Fig3a}
\end{subfigure}\hfill
\begin{subfigure}{.5\textwidth}
     \includegraphics[width=\linewidth]{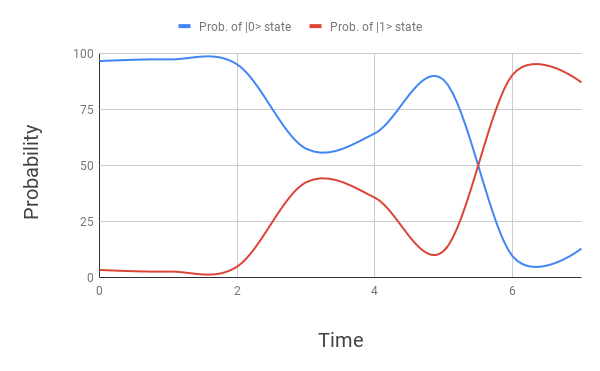}
    \caption{ For $\omega$=2}
    \label{qho_Fig3b}
\end{subfigure}\hfill

\begin{subfigure}{.5\textwidth}
      \includegraphics[width=\linewidth]{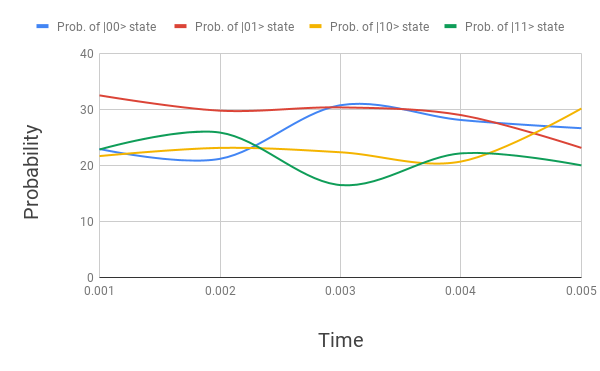}
    \caption{For $\omega$=1}
    \label{qho_Fig3c}
\end{subfigure}\hfill
\begin{subfigure}{.5\textwidth}
    \centering
     \includegraphics[width=\linewidth]{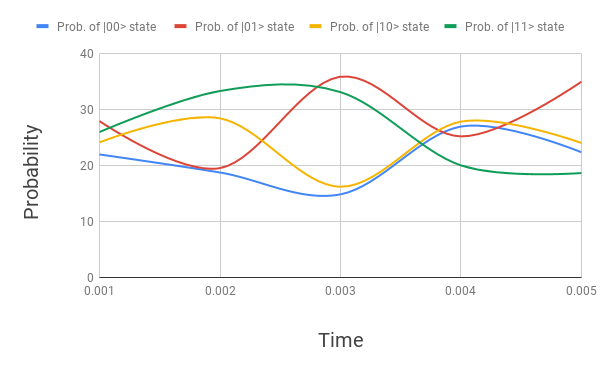}
    \caption{For $\omega$=5}
    \label{qho_Fig3d}
\end{subfigure}\hfill
\caption{Above Fig. \ref{qho_Fig3a}, \ref{qho_Fig3b}, \ref{qho_Fig3c} and \ref{qho_Fig3d} are the experimental run results for the time dependent quantum harmonic oscillator. Figs. \ref{qho_Fig3a} and \ref{qho_Fig3b} are for single-qubit case where the circuit was run for $\omega$ = 1 and 2 unit. Again, considering $\omega$ as 2 and 5 units for two-qubit system, the run results are plotted in SubFigs \ref{qho_Fig3c} and \ref{qho_Fig3d}. We have taken 1024 shots in the IBM quantum computers to run all the quantum circuits.}
\label{qho_Fig3}
\end{figure*}

We have studied the behavior of a particle in QHO by varying the oscillation frequency of the periodic force $\omega$. In Figs. \ref{qho_Fig3a} and \ref{qho_Fig3b}, the blue line indicates the probability distribution for the ground state and the red line for the first excited state. For both the cases, initially the particle has a higher probability, and with the evolution of time, the particle shows a fluctuating probability distribution for both the states. For Fig. \ref{qho_Fig3a}, we have considered $\omega$=1 unit and time in the range of 1 to 5 unit. In Fig. \ref{qho_Fig3b}, $\omega$ was taken as 2 unit, and time range is the same as above. Relevant data are available in Table \ref{qho_Table2}, for $\omega$ = 1  and $\omega$ = 2 in single qubit case respectively. \\

In Figs. \ref{qho_Fig3c} and \ref{qho_Fig3d}, the probabilities for ground state, first excited state, second excited state and third excited state are shown very clearly. In Fig. \ref{qho_Fig3c}, considering $\omega$ as 1, in the ground state, initially the probability decreased up to a certain time after then the probability is increased, again the probability drops slowly. Similarly, we can explain the behavior of the particle for other states also. Again, in Fig. \ref{qho_Fig3d}, the probability distribution of the particle is plotted considering $\omega$ as 5 unit. If we see Fig. \ref{qho_Fig3d}, the fluctuation of the probability distribution is higher than the case of $\omega$=1 unit. The relevant data for two-qubit case are available in Tables \ref{qho_Tab3} and \ref{qho_Tab4}. All the quantum circuits are run on the real quantum chip 'ibmqx4', for 1024 shots. 

\begin{table}[H]
       \centering
      
       \begin{tabular}{c | c c | c c}
       
       \hline
       \hline
        Time & $\Ket{0}$ & $\Ket{1}$ & $\Ket{0}$ & $\Ket{1}$\\
       \hline
       \hline
       1 & 98.45 & 1.55 & 96.58 & 3.42 \\
       2 & 95.91 & 4.09 & 97.35 & 2.65 \\
       3 & 90.61 & 9.39 & 95.11 & 4.89 \\
       4 & 42.74 & 57.26 & 57.51 & 42.49\\
       5 & 40.62 & 59.38 & 64.36 & 35.64\\
       6 & 25.88 & 74.12 & 88.14 & 11.86\\
       7 & 15.61 & 84.39 & 9.52 & 90.48\\
        \hline
        \hline
       \end{tabular}
       \caption{Probability (in percentage) considering $\omega=1$ (2nd column) and $\omega$=2 (3rd column) for single qubit case.}
       \label{qho_Table2}
\end{table}

\begin{table}[H]
        \centering
        \begin{tabular}{c c c c c c}
        \hline
        \hline
        Time  & $\Ket{00}$ & $\Ket{01}$ & $\Ket{10}$ & $\Ket{11}$\\
        \hline \hline
        1 & 22.949 & 32.52 & 21.68 & 22.852\\
        2 & 21.191 & 29.785 & 23.145 & 25.879\\
        3 & 30.762 & 30.371 & 22.363 & 16.504\\
        4 & 28.125 & 29.004 & 20.705 & 22.168\\
        5 & 26.66 & 23.145 & 30.176 & 20.02\\
        \hline
        \hline
       \end{tabular}
       \caption{Probability (in percentage) for $\omega=1$ for two qubits case}
       \label{qho_Tab3}
\end{table}

\begin{table}[H]
       \centering
       \begin{tabular}{c c c c c c}
       \hline
       \hline
        Time & $\Ket{00}$ & $\Ket{01}$ & $\Ket{10}$ & $\Ket{11}$\\
        \hline \hline
        1 &  21.973 & 27.93 & 24.121 & 25.977\\
        2 & 18.75 & 19.531 & 28.418 & 33.301\\
        3 & 14.844 & 35.84 & 16.211 & 33.105\\
        4 & 26.953 & 25.195 & 27.832 & 20.02\\
        5 & 22.363 & 34.961 & 24.023 & 18.652\\
        \hline
        \hline
       \end{tabular}
       \caption{Probability (in percentage) for $\omega=5$ for two-qubit case.}
       \label{qho_Tab4}
        \end{table}
        
\section{Conclusion \label{qho_conclusion}}
In this work, we have done the quantum simulation of the quantum harmonic oscillator by using the IBM Q experience platform. We have found the results for both the single-qubit and two-qubit cases. For the single-qubit case, we have taken the states of the particle to be $\Ket{0}$ and $\Ket{1}$. In case of two qubits, four states have been considered; those are $\Ket{00}$, $\Ket{01}$, $\Ket{10}$ and $\Ket{11}$ states. The probabilities of finding the states have been performed with different frequencies, and the graphs have been plotted. All the physical systems that we come across can be mapped to a two-level system. The quantum harmonic oscillator is a well-known example of a two-level system. Hence, quantum simulation of many physical systems can be understood from the simulation of the quantum harmonic oscillator.\\

\section*{Acknowledgements} A.B. and L.K. acknowledge the hospitality of Indian Institute of Science Education and Research Kolkata during the project work. B.K.B. acknowledges the financial support of IISER Kolkata. We acknowledge the support of IBM Quantum Experience for using the quantum processors. The views expressed are those of the authors and do not reflect the official position of IBM or the IBM quantum experience team. 

\end{document}